\documentclass[aip,twocolumn]{revtex4}
\usepackage{graphics}
\usepackage{epsfig}
\usepackage{bm}

\begin{document}

\title{Signature of a universal statistical description for drift-wave plasma turbulence}

\author{Johan Anderson and Pavlos
Xanthopoulos \\
Max-Planck-Institut f\"{u}r Plasmaphysik, IPP-Euratom Association, \\
Teilinstitut Greifswald, D-17491 Greifswald, Germany}

\begin{abstract}
\noindent
 This Letter provides a theoretical interpretation of numerically
 generated probability density functions (PDFs) of intermittent plasma
 transport events. Specifically, nonlinear gyrokinetic simulations of
 ion-temperature-gradient turbulence produce time series
 of heat flux which exhibit manifestly non-Gaussian PDFs with enhanced tails. It is demonstrated
 that, after the removal of autocorrelations, the numerical PDFs can be 
 matched with predictions from a fluid theoretical setup,
 based on the instanton method. This result
 points to a universality in the modeling of intermittent stochastic process,
 offering predictive capability.
\end{abstract}

\maketitle

Stochastic physical processes are most often observed to be unimodal
with exponential tails~\cite{a10}, a feature which is also attributed to
fluctuations in magnetically confined plasmas~\cite{a11}-~\cite{a16}.
These fluctuations are intermittent events manifesting a patchy spatial
and bursty temporal structure, pertaining to radially
propagating coherent structures like blobs or avaloids \cite{antar}, and have been suggested to carry
a significant fraction of the total transport \cite{car}.
Therefore, a comprehensive
predictive theory is called for, in order to understand and,
subsequently, improve properties related to intermittency. A major goal
would be, for instance, the control of edge heat flux loads, which depend on the instant 
amplitude of fluctuations, as opposed to the mean load,
which can be calculated by quasilinear theory.

In terms of mathematical description, the likelihood of intermittent
events related to plasma turbulence is expressed by probability density functions (PDFs), which
usually deviate significantly from the Gaussian distribution. 
Along these lines, there have been attempts to characterize the
statistical properties of the PDFs, based on phenomenological  
premises \cite{a28,krom} or numerical
investigations \cite{sanch,dif} alone. Here, however, 
we carry out a direct comparison between
first-principles analytical modeling and numerical simulations. 
As will be evident in the sequel, although the two approaches express
the same physics, they nevertheless greatly differ in their theoretical backgrounds. 
A key finding of this work is that the intermittent
process in the context of drift-wave turbulence appears to be independent of the specific modeling
framework, opening the way to the prediction of its salient features. 

The main part of this Letter consists in providing a theoretical interpretation of the
PDFs of radial heat flux derived by nonlinear, local, gyrokinetic (gk)
simulations of drift wave turbulence in tokamaks. Our paradigm in this work is 
Ion-Temperature-Gradient (ITG) turbulence with adiabatic
electrons~\cite{a17}. The simulations have been carried out with the
{\sc GENE} code~\cite{gene} in
a simple large aspect ratio, circular tokamak geometry. In particular,
we calculate the turbulent ion radial heat flux, $Q=\left <{v}_r \int \Big(\frac{1}{2} m_i
v_\|^2+ \mu B \Big) F_i \right >$, where ${v}_r$
is the radial $\bm{E} \times \bm{B}$ velocity, $v_\|$ the parallel
velocity,
$\mu$ the magnetic moment, $m_i$ the ion mass, $F_i$
the perturbed ion gyrocenter distribution function and $B$ the modulus
of the magnetic field. The brackets denote spatial averaging over the entire simulation
domain.
In order to perform a reliable comparison, we produce a
series of different cases, by varying the magnetic shear
$\hat{s}=\frac{r}{q} \frac{dq}{dr}$ ($q$ is the safety factor, i.e., the
number of toroidal turns of the magnetic field for each polodal turn), which appears both in
the gk simulations and the theoretical model. There is nothing special
about the selection of this parameter, and a similar scan could be
performed over, for example, the ion temperature or density gradient. Here, we
have set the (normalized) ion temperature gradient $R/L_{T_i} = 9$,
the density gradient $R/L_n = 2$, and the ion-to-electron temperature ratio  $\tau =1$.

For each case, the time evolution of the ion heat flux is considered as a time series, to which we
apply standard Box-Jenkins modeling~\cite{box}. This mathematical procedure
effectively removes deterministic autocorrelations
from the system, allowing for the statistical
interpretation of the residual part, which a posteriori turns out to be relevant
for comparison with the analytical theory. In our setup, it turns out
that an ARIMA(3,1,0) model accurately describes the stochastic
procedure, in that, one can express the (differenced) heat flux time trace in the form
\begin{eqnarray}
Q_{t+1}=a_1\,Q_t+a_2\,Q_{t-1}+a_3\,Q_{t-2}+Qres(t)
\end{eqnarray}
where the fitted coefficients $a_1, a_2, a_3$ decribe the deterministic component
and $w(t)$ is the residual part (noise). The reliability of the
produced autoregressive model was routinely verified via portmanteau testing and overfitting. 

\begin{figure}[ht]
  \centering
{\centering \resizebox*{0.7\columnwidth}{6cm}{\includegraphics{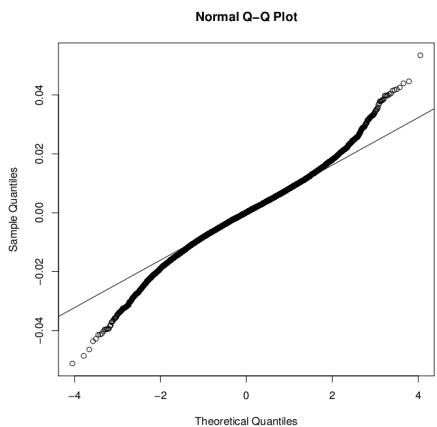}} \par}
  \caption{Normal quantile-quantile plot showing strong deviation of the residual PDF
 from Gaussian statistics, in view of the enhanced tails representing about half of the
 sample.}
  \label{figure1}
\end{figure}

It is systematically observed that the residual PDFs are manifestly
non-Gaussian with elevated tails, as shown for instance in Fig.~\ref{figure1}, for the case  
$\hat{s}=1.0$, where the sample residuals from a {\sc GENE} simulation are
tested against the Gaussian distribution, via a normal quantile-quantile
plot. 
Furthermore, two significant statistical quantities, namely the variance
$\sigma^2=\left<Q^2 \right>-\left<Q\right>^2$ and the kurtosis
$K=\left<Q^4\right>/\sigma^4$ (here, the brackets denote averaging over
the statistical sample), during the magnetic shear scan are
summarized in Table~1. Evidently, the decrease of the variance and the
rapid increase in kurtosis for increasing magnetic shear renders a
Gaussian description improper. We note in passing, that the values of
$\hat{s}$ were selected such that the resolution of the simulation
domain remains constant, thus avoiding numerical artifacts.

In the sequel, it will be shown that the aforementioned PDF tails can be satisfactorily
predicted, starting from a fluid model consisting of a continuity and an energy
equation~\cite{joh}, and using a nonperturbative statistical technique,
called the instanton method, which has been
adopted from Quantum Field Theory and then modified to classical
statistical physics for Burgers turbulence and in the passive scalar
model of Kraichnan~\cite{balk}. In this context, the key element is to
identify the bursty or intermittent event with the appearance of a
coherent structure (e.g., streamer).

\begin{table}[ht]
\caption{Simulation Results} 
\centering 
\begin{tabular}{c c c c c c c} 
\hline\hline 
$\hat{s}$ & Variance & Kurtosis \\ [0.5ex] 

 0.25 & 5.962e-4 & 3.529 \\ 
 0.40 & 5.253e-4 & 3.517 \\
 0.50 & 6.049e-4 & 3.527 \\
 0.60 & 5.874e-4 & 3.715 \\ 
 0.75 & 4.403e-4 & 4.324\\ 
 1.00 & 8.253e-5 & 14.413 \\ 

\end{tabular}
\label{table:simulat} 
\end{table}

In the following paragraph, we briefly outline the implementation of the instanton method. For
more details, the reader is referred to the existing literature ~\cite{a18}-~\cite{a22}.
The PDF tail is first formally expressed in terms of a path integral by
utilizing the Gaussian statistics of the forcing in the continuity
equation in a similar spirit as in~\cite{car}. An optimum path will then be associated with the creation of a
modon (among all possible paths or functional values) and the action
($S_{\lambda}$, below) is evaluated using the saddle-point method on the
effective action in the limit $\lambda \rightarrow \infty$. The
instanton is localized in time, existing during the formation of the
modon. The saddle-point solution of the dynamical variable
$\phi(\vec{x},t)$ of the form $\phi(\vec{x},t) = F(t) \psi(\vec{x})$ is
called an instanton if $F(t) = 0$ at $t=-\infty$ and $F(t) \neq 0$ at
$t=0$. Note that, the function $\psi(\vec{x})$ here represents the
spatial form of the coherent structure. Thus, the intermittent character
of the transport consisting of bursty events can be described by the creation of modons.
The probability density function of the heat flux $Q$ can be defined as 
\begin{eqnarray} \label{pq}
P(Q) =  \langle \delta(v_r n T_i(\vec{r}=\vec{x}_0)) - Q) \rangle = \int d \lambda e^{i \lambda Q} I_{\lambda},
\end{eqnarray}
where 
\begin{eqnarray} \label{ilambda1}
I_{\lambda} = \langle \exp(-i \lambda v_r n T_i(\vec{x}=\vec{x}_0)) \rangle.
\end{eqnarray}
Here, $v_{r}$ is the radial drift velocity and $T_i$ the ion temperature. The integrand can then be rewritten in the form of a path-integral as
\begin{eqnarray} \label{ilambda2}
I_{\lambda} = \int \mathcal{D} \phi \mathcal{D} \bar{\phi} e^{-S_{\lambda}}.
\end{eqnarray}
Although $\bar{\phi}$ appears to be simply a convenient mathematical
tool, it does have a useful physical meaning that should be noted; it
arises from the uncertainty in the value of $\phi$ due to the stochastic
forcing. That is, the dynamical system with a stochastic forcing should
be extended to a larger space involving this conjugate variable, whereby
$\phi$ and $\bar{\phi}$ constitute an uncertainty relation. Furthermore,
$\bar{\phi}$ acts as a mediator between the observables (heat flux) and
instantons (physical variables) through stochastic forcing. In Eq.~\ref{ilambda2},
the integral in $\lambda$ is computed using the saddle-point method
where it is shown that the limit $\lambda \rightarrow \infty$
corresponds to $Q \rightarrow \infty$, representing the tail part of the distribution.
Based on the assumption that the total PDF can be characterized by an
exponential form, the expression
\begin{eqnarray} \label{pq2}
P(Q) & = & \frac{1}{Nb} \exp{\{ - b |Q-\mu |^{3/2}\}}, \\ \label{b}
b & = & b_0 (\frac{R}{L_n} + 2 \langle g_i \rangle \beta - U - \langle k_{\perp}^2\rangle (U + \frac{R}{L_n})), \\
\beta & = & 2 + \frac{2}{3} \frac{R/L_n - U}{U + 10/3 \tau \langle g_i \rangle}.
\end{eqnarray}
is found~\cite{a22}, where the heat flux $Q$ plays the role of the stochastic
variable, with $P(Q)$ determining its statistical properties.
Several auxiliary definitions are also utilized; the normalization constant $N$; the gradient scale
lengths $L_f = - \left( d ln f / dr\right)^{-1}$; the normalized modon speed ${U} =
R U/L_n$ and temperature ratio $\tau = T_i/T_e$; $g_i = \omega_d/\omega_{\star} = \frac{2 L_n}{R} (\cos (\xi) + \hat{s} \xi \sin(\xi))$ where $\omega_d$ is the curvature drift frequency and $\omega_{\star}$ is the diamagnetic drift frequency; $k_{\perp}^{2} = k_y^2 (1 + \hat{s}^2 \xi^2)$ is the perpendicular wave number; the brackets denote averaging along the field line, e.g. for an arbitrary scalar function $f$, $\langle f \rangle = \int_{- \pi}^{\pi} d\xi \phi f \phi/ \int_{-
\pi}^{\pi} d \xi \phi^2$ where the eigenfunctions $\phi (\xi)$ are
extracted from the {\sc GENE} simulations and $\mu=\left<Q\right>$ is the mean value of the heat flux. The coefficient $b_0$ is a free parameter and represents the strength of the forcing in the continuity equation. Note that the proposed PDF is close enough to a Gaussian
distribution to match the bulk of the PDF while retaining the fat
tails. Furthermore, the exponential form of the PDF will be the same for
momentum flux with a modified coefficient $b$, which is in agreement
with findings in recent experiments~\cite{a24}.
 
\begin{figure}[h]
  \centering
{\centering \resizebox*{0.7\columnwidth}{6cm}{\includegraphics{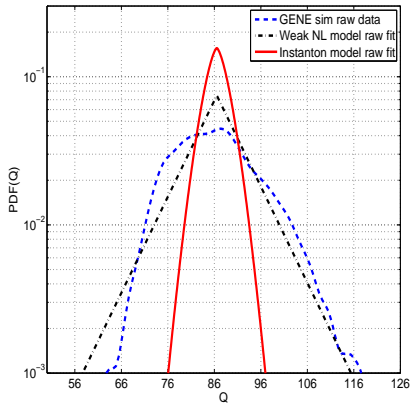}} \par}
  \caption{The numerically estimated PDF of heat flux (dashed blue line) in comparison with
 the analytically predicted instanton method (red line) and the weak non-linear model (dashed-dotted black line) at magnetic shear $\hat{s} = 0.25$.}
  \label{figure2}
\end{figure}

In Fig.~\ref{figure2} ($\hat{s} = 0.25$), the PDF of heat flux $Q$ stemming
from the {\em raw} timetrace of the simulation (blue line) is shown. Although a simple visual
inspection precludes the Gaussian distribution, the analytically predicted
instanton method from Eq.~\ref{pq2} (red line) does a rather poor job
modeling the simulation result. This is, however, contrasted to the PDF using the weak non-linear model (dashed-dotted black line) where reasonably good agreement is found similar to what have been reported earlier~\cite{a15, car}. The weak non-linear model (a Laplace distribution) is also predicted by the instanton model~\cite{kim} when the non-linearities are neglected and has the form $P(Q) = e^{-|Q-\mu|/b}/(2b)$ where $b$ is determined by the variance. The coefficient $b_0$ in Eq.~\ref{b} is
determined from the simulations at the point $\hat{s} = 0.5$. 

\begin{figure}[h]
  \centering
{\centering \resizebox*{0.7\columnwidth}{6cm}{\includegraphics{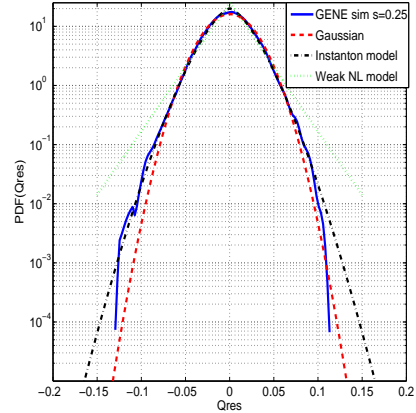}} \par}
  \caption{Comparison of the numerically estimated PDF (blue line) against the
 analytically predicted instanton model (dashed-dotted black line) and weak non-linear model (dashed green line) compared to the best fit Gaussian (dashed red line) for $\hat{s} = 0.25$.}
  \label{figure3}
\end{figure}

\begin{figure}[h]
  \centering
{\centering \resizebox*{0.7\columnwidth}{6cm}{\includegraphics{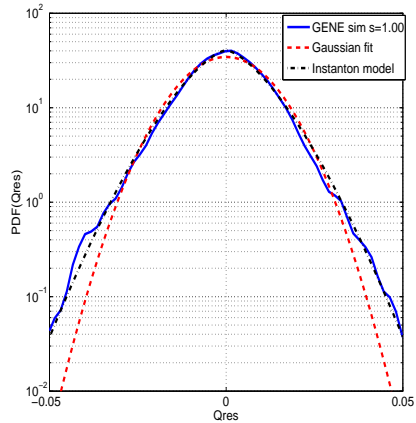}} \par}
  \caption{Comparison of the numerically estimated PDF (blue line) against the
 analytically prediction (dashed dotted black line) and the best fit Gaussian (dashed red line) for $\hat{s} = 1.0$.}
  \label{figure4}
\end{figure}

Figs.~\ref{figure3} and \ref{figure4} single out the comparison of the
 numerically estimated residual PDFs against the instanton prediction in Fig.~\ref{figure1}.
 Here, the PDF of $Qres$ from the simulation (blue line) and the analytically
 predicted PDF from the instanton method (dashed-dotted black
 line) demonstrate a significantly better agreement both at the tails
 and the center of the PDF, as compared to the best fit (using the first two moments)
 Gaussian distribution (dashed red line) at magnetic shears $\hat{s} = 0.25$
 and $\hat{s} = 1.0$, respectively. Note, we show the weak non-linear model with a value of the variance of 8.00e-4 for better agreement and that the distribution does not capture the form of the stochastic PDF.
The surprisingly good agreement between the gyrokinetic and fluid descriptions of
 drift wave turbulence strongly suggests a universality
 of the statistics of the intermittent process.

\begin{figure}[h]
  \centering
{\centering \resizebox*{0.7\columnwidth}{6cm}{\includegraphics{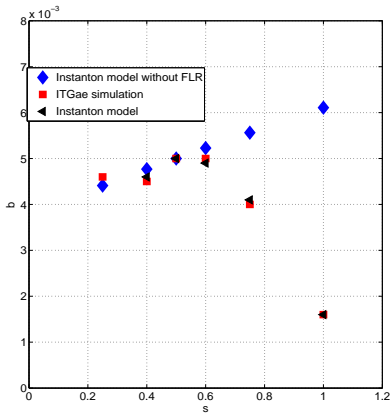}} \par}
  \caption{The coefficient $b$ as a function of magnetics shear
 $\hat{s}$. The simulation results (red squares) are compared to the
 analytical fluid model prediction (black stars) and the same with
 suppressed FLR effects (blue diamonds).}
  \label{figure5}
\end{figure}

Since the quality of the agreement in the previous figures relies
heavily on the exponential coefficient $b$ in Eq. 5, this is displayed for the whole magnetic shear scan
in Fig.~\ref{figure5}, verifying that good agreement is expected for any
value of $\hat{s}$. Here, we note that the abrupt bending of the curve
for larger values of shear is solely attributed to FLR effects.

In conclusion, we have presented a first quantitative paradigm of
universality in intermittent stochastic processes related to drift wave
turbulence. Numerical PDFs of heat flux were generated with the gyrokinetic code GENE
in the framework of toroidal ITG turbulence, and subsequently processed
with Box-Jenkins modeling, in order to remove deterministic
autocorrelations, thus retaining their stochastic parts only. These PDFs have been
shown to agree very well with analytical predictions based on a fluid
model, on applying the instanton method. Specifically, we were able to quantitatively
confirm the exponential form of the PDFs, therefore adding the important element of
predictive strength to the existing phenomenological approaches.

In future publications we will address the emergent universal scalings
of the PDFs of potential, density and temperature where the theory predicts different tails~\cite{kim}. A study of potential or density fluctuations opens up opportunities for comparison with experimentally measured PDFs. Still within the scope of ITG turbulence, we will include effects from kinetic electrons, in order to test the robustness of the exponential scaling. A quite interesting complementary work would involve incorporating results from nonlinear fluid simulations, based on Braginskii-like equations, and relate them to the already
presented findings. Finally, our present setup provides a
testbed for the investigation of the impact of zonal flows on the
statistics, as reported elsewhere~\cite{sanch}. Work addressing these
avenues has been initiated.

\section*{Acknowledgments} The GENE simulations have been performed
at the J\"ulich Supercomputing Center (JSC). The Authors are grateful 
to Prof. P. Helander and Prof. F. Jenko for many fruitful discussions.

\end{document}